\documentclass[12pt]{article}
\usepackage{amstext}
\usepackage{amssymb}
\newtheorem{defi}{Definition}[section]
\newtheorem{lem}[defi]{Lemma}
\newtheorem{prop}[defi]{Proposition}
\newtheorem{theorem}[defi]{Theorem}
\newtheorem{cor}[defi]{Corollary}

\def\bt{\begin{theorem}}
\def\et{\end{theorem}}
\def\le{\left}
\def\ri{\right}
\newcommand{\bfrakg}{\mbox{\boldmath $\mathfrak g$}}
\newcommand{\bfrakh}{\mbox{\boldmath $\mathfrak h$}}
\newcommand{\bfrakX}{\mbox{\boldmath $\mathfrak X$}}

\newcommand{\h}{\mbox{\boldmath $\mathfrak h$}}
\def\R{{\mathbb R}}
\def\C{{\mathbb C}}
\def\O{{\mathcal O}}
\def\d{{\mathrm d}}

\def\G{\R^n\rtimes H}

\def\Sinch{{\, \hbox{sinch}\,}}
\def\sinh{{\, \hbox{sinh}\,}}
\def\exp{{\, \hbox{exp}\,}}

\def\be{\begin{equation}}
\def\ee{\end{equation}}
\def\en{\end{equation}}
\def\bea{\begin{eqnarray}}
\def\eea{\end{eqnarray}}

\newcommand{\bedefin}{\begin{defi}}

\newcommand{\betheo}{\begin{theorem}}
\newcommand{\enth}{\end{theorem}}
\newcommand{\entheo}{\end{theorem}}

\newcommand{\becor}{\begin{cor}}
\newcommand{\encor}{\end{cor}}

\newcommand{\belem}{\begin{lem}}
\newcommand{\enlem}{\end{lem}}

\newcommand{\beprop}{\begin{prop}}
\newcommand{\enprop}{\end{prop}}

\newcommand{\beano}{\begin{eqnarray*}}
\newcommand{\enano}{\end{eqnarray*}}

\newcommand{\bee}{\begin{enumerate}}
\newcommand{\ene}{\end{enumerate}}

\newcommand{\bei}{\begin{itemize}}
\newcommand{\eni}{\end{itemize}}

\newcommand{\betab}{\begin{tabular}}
\newcommand{\entab}{\end{tabular}}

\newcommand{\hil}{\mbox{${\mathfrak   H}$}}

\newcommand{\bTheta}{\mbox{\boldmath $\Theta$}}

\newcommand{\bh}{\mathbf h}

\newcommand{\bT}{\mathbf T}

\def\pr{^\prime}

\def\Ad#1{\mathop{{\rm Ad}_{{#1}}}}
\def\coAd#1{\mathop{{\rm Ad}_{{#1}}\dual}}
\def\dual{^{\sharp}}

\def\lb{\hbox{$\lambda$\hskip-5pt$\raise2pt\hbox{\char'040}$\hskip1.5pt}}

\begin{document}
\baselineskip 30pt plus 2pt minus 2pt

\begin{center}
{\LARGE \bf  Wigner Functions for a Class of \\[4mm]
      Semidirect Product Groups\footnote{Based in part on a
thesis submitted by one of us (AEK) to Concordia University.}}\\[10mm]
\sc  A. E. Krasowska  {\rm and} S. Twareque Ali\\[2mm]
{\it Department of Mathematics and
Statistics\\ Concordia University, Montr\'eal\\
Qu\'ebec, Canada H4B 1R6}\\[10mm]
\end{center}
\begin{abstract}
\baselineskip 30pt plus 2pt minus 2pt
Following a general method proposed earlier, we construct here Wigner
functions defined on coadjoint orbits of a class of semidirect
product groups. The groups in question are such that their unitary
duals consist purely of representations from the discrete series and
each unitary irreducible representation is associated with a
coadjoint orbit. The set of all coadjoint orbits (hence UIRs) is
finite and their union is dense in the dual of the Lie algebra. The
simple structure of the groups and the orbits enables us to
compute the various quantities appearing in the definition of the
Wigner function explicitly. Possible use of the Wigner functions so
constructed, in image analysis and quantum optical measurements, is
suggested.
\end{abstract}


\newpage
\section{Introduction}\label{sec:intro}

 The Wigner function, a useful computational tool
in atomic and
quantum statistical physics, and more recently in quantum optics
and image processing, was introduced by Wigner \cite{Wig-1932} as
a {\em quasi-probability distribution} on phase space in the context
of quasi-classical approximations. Using
this distribution function, each quantum
mechanical state can be represented as a real-valued
(not necessarily positive) function of the position and momentum
variables $\vec q, \vec p$, of a free particle moving on
the classical (flat) phase space $\Gamma = {\mathbb R}^{2n}$.
In view of the wide
applicability of the Wigner function, and its deeper significance
in harmonic analysis \cite{AACW-1999,Bast-78,Bast-79,KBW-96},
it is useful to try to
construct analogous (quasi-probability) distribution functions
for more general phase spaces and quantum systems exhibiting
specific symmetries. Such an attempt was initiated in
some earlier work, where we developed a general procedure for
constructing Wigner functions on coadjoint orbits of certain
Lie groups \cite{AACW-1999,AKM-2000,AFK-2001}. As is
well-known from the
Kirillov theory \cite{Kirill-76}, coadjoint orbits of Lie
groups carry symplectic structures, equipped with
intrinsically defined Liouville type invariant
measures, making them resemble classical phase spaces.
In particular, for semi-direct product groups of the
type $G = {\mathbb R}^{n}\rtimes H$, where $H$ is a
closed subgroup of
$GL (n, {\mathbb R})$, the coadjoint orbits of interest to
us for the construction of Wigner functions,
are cotangent bundles \cite{AAG-2000,GS},
$T^{*}{\widehat \O}$, of manifolds ${\widehat \O}$ which are
themselves orbits, under the action of $H$, of vectors in the
the dual space ${\widehat{\mathbb R}}^{n}$ (of
the vector space ${\mathbb R}^{n}$).  The
construction of Wigner functions outlined in \cite{AACW-1999},
and extended in \cite{AFK-2001},
works particularly well for such groups and especially when
the orbit ${\widehat \O}$ is open and free. This means that these
orbits have dimension $n$, are open sets of
${\widehat{\mathbb R}^{n}}$ and
the stability subgroup consists of just the identity element $e$
of $H$. The unitary irreducible representations of these groups
have the property of being {\em square integrable}, an
aspect studied in detail in \cite{BT-96}.
In this paper we shall explicitly construct Wigner functions
for a number of such groups, which are  of importance in
image processing, signal analysis and quantum optics. We also derive
expressions for the generic group $G = {\mathbb R}^{n}\rtimes H$. The
advantage of working with such groups is that the various quantities
which appear in the definition of the Wigner function can be
explicitly calculated. For several of the examples worked out here,
Wigner functions have also been constructed in the past using
special techniques, specifically adapted to each group being examined
\cite{BB-92,BB-84}. However, ours is a general technique, applicable
to a wide class of Lie groups.
\par
The rest of this paper is organized as follows: Section 2 is a brief 
recapitulation of the standard Wigner function and its properties; Section 3 
lays out some preliminary mathematical properties of the kind of semidirect 
product groups, for which we shall be constructing Wigner functions in 
this paper. In Section 4 we describe, in some detail, the symplectic structure 
of the coadjoint orbits of the groups in question and in Section 5 we describe 
their associated unitary irreducible representations. Section 6 recapitulates the 
notion of a square integrable representation and sets out some results on 
the square integrability of the irreducible representations introduced in the 
previous section. Sections 7, 8 and 9 lay down the main results of this 
paper, by first introducing the general construction of Wigner functions 
for square integrable representations, followed by their properties and 
finally, the specialization 
of the construction to semidirect product groups with open free orbits. 
Section 10 is devoted to the question of the support of the Wigner functions 
constructed in the previous section. In Section 11 a large number of explicit 
examples are worked out. 
\section{The standard Wigner function}\label{sec:stanwigfcn}
 Let us begin with a quick revision of  some basic
properties of the function defined in \cite{Wig-1932}.
The  quasiprobability distribution function $W^{QM}$ is defined,
on the flat phase space $\Gamma
=\R^{2n}$, for  any quantum mechanical state $\phi \in
L^2 (\R^n , \d\vec x )$ as
\be
  W^{QM}(\phi\;\vert\; \vec q ,\vec p;h) = \frac 1{h^n} \int_{\R^n}
  \overline{\phi(\vec q - \frac {\vec x}2)}\;
 e^{-\frac{2\pi i \vec x \cdot  \vec p}{h}} \;
\phi(\vec q + \frac {\vec x}2) \;\d \vec x \;.
\label{1}
\ee
The variable $ \vec q $ represents the position of
the system,
$\vec p$ its momentum at the point $\vec q$ and $h$ is
Planck's constant. The phase space
$\Gamma $ can also be viewed as an orbit $\O^{*}$ under the
coadjoint action of the Heisenberg-Weil group $ G_{HW} $ on the
dual space
$\bfrakg_{HW} ^{*}$ of its Lie algebra $\bfrakg _{HW}$, and this
is the point of departure in \cite{AACW-1999} for constructing a
generalization of the Wigner function.
Canonical transformations of the phase space $\Gamma$:
\be
(\vec q, \vec p) \to (\vec q - \vec q_0 ,\vec p - \vec p_0 ),
\label{2}
\ee
which equivalently can be viewed as the coadjoint action
of $ G_{HW}$
on $\bfrakg _{HW}^*$,  lead to  unitary transformations
on the space of wave functions
$\phi$:
\be
\phi \longrightarrow U(\vec q_0, \vec p_0) \phi =
  e^{\frac{2 \pi i}{h}(\vec Q\cdot \vec p_0 -
           \vec P\cdot\vec q_0)} \phi ,
\label{3}
\ee
where $\vec Q $ and $\vec P$ are the usual n-vector
operators of position and momentum respectively.
It can be shown that the Wigner function satisfies the following
covariance condition related to (\ref{2}) and (\ref{3}):
\be
 W^{QM}(U(\vec q_0,\vec p_0) \phi \;|\; \vec q,\vec p ;h)=
       W^{QM}(\phi \;|\; \vec q - \vec q_0,\vec p - \vec p_0;h)
\label{covariance}
\ee
The other important property of the Wigner function is
the existence of the marginaliy conditions, which make it resemble
a classical probability distribution:
\be
\int_{\R^{2n}} W^{QM}(\phi \;|\;\vec q,\vec p ; h)\;\d \vec p
          = |\phi (\vec q)|^2 \; , 
\label{marginal1}
\ee
and
\be
\int_{\R^{2n}} W^{QM}(\phi \;|\; \vec q,\vec p ;h)\;\d \vec q =
        |\widehat \phi (\vec p)|^2 \; ,
\label{marginal2}
\ee
where $\widehat \phi$ is the Fourier transform of $\phi$. However,
as expected, for a given $\phi$ there exist in general regions of
phase space over which the function $W^{QM}(\phi \;|\;\vec q,\vec p , h)$
can also assume negative values and hence $W^{QM}$ cannot
be a true probability density. An obvious generalization leads to a 
definition of the Wigner function for a pair of wave functions $\phi$,
$\psi$; one then has the well known 
{\em overlap condition\/}:
\be
\int_{\R^2}\d \vec q \d \vec p W^{QM}(\phi, \psi|\vec q,\vec p)
         W^{QM}(v,w|\vec q,\vec p)=<\phi|w><v|\psi>\; .
\label{overlap1}
\ee
   A general Wigner function, defined using some other group,
should also preserve as many of the above properties as possible.
Moreover, mathematically it should appear in much the same way,
as a function defined on a coadjoint orbit. The work reported in
\cite{AACW-1999,AKM-2000} was an attempt to do this.
\section{Some mathematical preliminaries}\label{sec:mathprelims}

We begin with some basic properties of semi-direct
product groups of the type mentioned above and in particular take
a closer look at their non-trivial orbits, i.e. orbits of maximal dimension.
Let  $G= \R^n\rtimes H$ be the semidirect product group with
elements $ g=(\vec b,{\bh})$, $\vec b \in \R ^n $, and
${\bh} \in H $ and the multiplication law :
\be
(\vec b_1,{\bh}_1)(\vec b_2,{\bh}_2)=(\vec b_1+{\bh}_1\vec
b_2,{\bh}_1{\bh}_2)
\label{groupmult}
\ee
Here $H$ is assumed to be a closed subgroup of $GL(n,\R)$ and,
as mentioned earlier, we will consider only the
case where $H$ is an n-dimensional subgroup of $GL(n,\R )$ such
that there exists at least one open free orbit,
 $\widehat\O_{\vec k^{\;T}} = \{{\vec k}^{\;T} {\bh}\; \vert\; {\bh} \in H \}$,
for some ${\vec k}^{\;T}$ in ${\widehat \R}^n$(the dual
of $\R^n$). (Our convention is to use column vectors for elements of 
$\R^n$ and row vectors for those of $\widehat{\R}^n$, the superscipt $T$ denoting 
a transpose.)
An element $g \in G$ can be written in matrix form as
$$ g=\pmatrix {{\bh}&\vec b\cr {\vec 0}^{\;T}&1} , $$
where ${\vec 0}^{\;T}$ is the zero
vector in $\widehat\R^n$. The inverse element is:
$$g^{-1}=\pmatrix {{\bh}^{-1}&-{\bh}^{-1}\vec b\cr {\vec 0}^{\;T}&1}.$$
Note that $\h$ is an $n\times n$ matrix with non-zero
determinant, which acts on
$\vec x \in {\mathbb R}^{n}$ from the left in the usual way,
$\vec x \mapsto \bh \vec x$, and similarly,  it acts on
$\vec x^{\;T}  \in \widehat{\mathbb R}^n$ from the right,
$\vec x^{\;T} \mapsto \vec x^{\;T}\bh$.
The left invariant Haar measure, $d\mu_{G}$, of $G$ is
\be
  d\mu_{G}(\vec b , \bh ) = \frac 1{\vert{\rm det}\;\bh\vert}\;
               d\vec b\; d\mu_{H}(\bh),
\label{lefthaar1}
\en
$d\vec b$ being the Lebesgue measure on $\R^n$ and $d\mu_{H}$
the left invariant Haar measure of $H$.
While it is the left invariant measure that we shall
consistently use, it is nonetheless worthwhile
to write down at this point the right invariant Haar measure
$d\mu_r$ as well, in terms of the left Haar measure and the modular
functions $\Delta_G ,\;\; \Delta_H$, of the groups $G$ and $H$,
respectively:
\be
  d\mu_{G}(\vec b , \bh ) = \Delta_{G}(\vec b , \bh )\;
       d\mu_{r}(\vec b , \bh ) =
     \frac {\Delta_{H}(\bh )}{\vert{\rm det}\;\bh\vert}\;
       d\mu_{r}(\vec b , \bh ).
\label{righthaar1}
\en

Let $\bfrakg = Lie(G)$ be the Lie algebra of $G$ and
$\{L^1,L^2,...,L^{2n}\}$ a basis of it chosen so that  
the first $n$ elements, $\{L^1,L^2,...,L^n\}$,
form  a basis of $\h = Lie(H)$, and the last $n$ elements,
$\{L^{n+1},L^{n+2},...,L^{2n}\}$, which are
the generators of translations, form a basis in $\R^n$.
An element $X \in \bfrakg$ can be written in matrix form as:
\be
  X=x_1L^1+x_2L^2+...+x_{2n}L^{2n}=
  \pmatrix{X_q& \vec x_p\cr {\vec 0}^{\;T}& 0}
\label{liealgelem1}
\ee
where $X_q$ is an $n \times n$ matrix with entries depending on
$x_i, \;\; i=1...n$,  and $\vec x_p$ is a column vector with
components $x_{n+1}, x_{n+2}, \ldots , x_{2n}$. Also it will
be useful to introduce the vector $\vec x_q$, with components
$x_i, \;\; i=1...n$, and the vector of matrices $\bfrakX = (L^1 ,
L^2 , \ldots , L^n )$. Next, for any $\vec u \in \R^n$, we
define the matrix $[\bfrakX \vec u]$ whose columns are the
vectors $L^i \vec u , \;\; i =1,2, \ldots n$,
\be
  [\bfrakX \vec u] = [L^1 \vec u ,\; L^2 \vec u  ,\;
             \ldots ,\;  L^n \vec u ].
\label{matvec}
\en
The adjoint action of the group on its Lie algebra is given by
\be
  X \mapsto \Ad{g}X := gXg^{-1} = \pmatrix{\bh X_q \bh^{-1}
           & -\bh X_q\bh^{-1}\vec b +
             \bh \vec x_p \cr
             {\vec 0}^{T} & 0}.
\label{adjaction1}
\en
Introducing the matrix $M(\bh )$ such that,
\be
     \bh L^k \bh^{-1} = \sum_{i=1}^{n}L^{i}M(\bh )_i^k ,
\label{adjaction2}
\en
the adjoint action of an element $g = (\vec b , \bh ) \in G$ may
conveniently be written in terms of its action on the $2n$-dimensional
vector $\pmatrix{\vec x_q \cr \vec x_p}$ as:
\be
   \pmatrix{\vec x_q \cr \vec x_p} \mapsto
   \pmatrix{\vec x_q^{\;'} \cr \vec x_p^{\;'}} =
    M(\vec b , \bh )\pmatrix{\vec x_q \cr \vec x_p},
\label{adjaction3}
\en
where $M(\vec b , \bh )$ is the $2n\times 2n$-matrix
\be
  M(\vec b , \bh ) = \pmatrix{M(\bh ) & {\mathbb O}_{n} \cr
                    -[\bfrakX\vec b]M(\bh ) & \bh},
\label{adjaction4}
\en
${\mathbb O}_{n}$ being the $n\times n$ null matrix. Note that
$M(\bh )$ is just the matrix of the adjoint action of $\bh$ on
$\bfrakh$ (the Lie algebra of $H$) computed with respect to the
basis $\{L^1 , L^2 , \ldots , L^n\}$. Similarly, $M(\vec b , \bh )$
is the matrix of the adjoint action of $g = (\vec b , \bh )$ on
$\bfrakg$, the Lie algebra of $G$. By abuse of notation, we shall
also write,
\be
   \Ad{h}X_q = M(\bh )\vec x_q , \qquad
    \Ad{g}X = M(\vec b ,\bh )\pmatrix{\vec x_q \cr \vec x_p }.
\label{adjaction5}
\en
The coadjoint action of $G$ on $\bfrakg^*$, the dual space
of its Lie algebra, can now be immediately read off from
(\ref{adjaction4}). Indeed, let $\{L^*_i\}_{i=1}^{2n}$ be the basis
of $\bfrakg^*$ which is dual to the basis $\{L^i\}_{i=1}^{2n}$ of
$\bfrakg$, i.e.,
$$  \langle L^*_i\; ; \; L^j \rangle = \delta_i^j ,
   \qquad i, j = 1,2, \ldots , 2n .$$
A general element $X^* \in \bfrakg^*$ then has the form,
$$ X^* = \sum_{i=1}^{n}\gamma^i L^*_i, \qquad \gamma^i \in \R $$
and again we introduce the row vectors,
$$\vec \gamma^{\;T} = (\gamma^1 , \gamma^2, \ldots,
\gamma^{2n} ), \;\; \vec \gamma_q^{\; T}  = (\gamma^1 , \gamma^2,
\ldots,  \gamma^n ), \;\; \vec \gamma_p^{\; T}  = (\gamma^{n+1} ,
\gamma^{n+2}, \ldots,  \gamma^{2n} ).$$
Using the relation
$$
  \langle \coAd{g}X^* \; ;\; X \rangle =
      \langle X^* \; ;\; \Ad{g^{-1}}X \rangle , $$
we easily obtain from (\ref{adjaction4})and with the same abuse of
      notation as in (\ref{adjaction5}),
\be
  \coAd{(\vec b , \bh)}X^* = (\vec \gamma_q^{\; T}, \;
     \vec \gamma_p^{\; T}) M( -\bh^{-1}\vec b , \bh^{-1})=
       (\vec \gamma_q^{\; T}, \; \vec \gamma_p^{\; T})
    \pmatrix{M(\bh^{-1} ) & {\mathbb O}_{n} \cr
                    \bh^{-1}[\bfrakX\vec b] & \bh^{-1}}.
\label{coadaction1}
\en
Thus, under the coadjoint action, a vector
$(\vec \gamma_q^{\; T}, \; \vec \gamma_p^{\; T})$ changes to:
\bea
  \vec \gamma_q^{\;'\; T} &= &\vec \gamma_q^{\; T}M(\bh^{-1})
        + \vec \gamma_p^{\; T}\bh^{-1}[\bfrakX\vec b], \nonumber\\
 \vec \gamma_p^{\;'\; T} &= &\vec \gamma_p^{\; T}\bh^{-1}.
\label{coadaction2}
\eea
Let us also note that the modular functions
appearing in (\ref{righthaar1}) can be written \cite{HG}
in terms of the coadjoint operators as:
\be
  \Delta_{G}(\vec b , \bh ) =
  \vert {\rm \det}\; \coAd{(\vec b , \bh)}\vert =
   \frac {\Delta_{H}(\bh )}
         {\vert {\rm \det}\;\bh \vert}, \qquad
\Delta_{H}(\bh ) =
  \vert {\rm \det}\; \coAd{\bh}\vert =
   \frac 1{\vert {\rm \det}\;M(\bh ) \vert}.
\label{modfcns}
\en
Before leaving this section we make a further
important assumption on the nature of the group $G$.
We require that the range in $G$ of the
exponential map be a dense set whose complement has
Haar measure zero. (This includes, for example, groups of
exponential type.) Thus, by exponentiating
(\ref{liealgelem1}),  we may
write any element (up to a set of measure zero) of $G$ as
\be
g=e^X=\pmatrix{e^{X_q}&e^{\frac {X_q}{2}}\Sinch \frac{X_q}{2}\ \vec
x_p\cr{\vec 0}^{T}&1}, \qquad X \in N ,
\label{expmap1}
\ee
where $N \in \bfrakg$ is the domain of the exponential map,
which contains the
origin and has the property that if $X \in N$ then $-X \in N$.
The  $n\times n$ matrix $\Sinch A$ is defined as the sum of an
infinite series:
\be
\Sinch A={\mathbb I}_n+\frac{1}{3!}A^2+
\frac{1}{5!}A^4+\frac{1}{7!}A^6+...
\label{sinchfcn1}
\ee
${\mathbb I}_n$ being the $n\times n$ unit matrix.
When the matrix  $A$ has an inverse, $\Sinch A $ can also
be formally written as:
\be
\Sinch A =\frac{e^A-e^{-A}}{2A} = A^{-1}\sinh A.
\label{sinchfcn2}
\ee
It will also be useful to introduce the matrix valued functions,
\bea
  F(X_q ) & = &e^{\frac {X_q}{2}}\Sinch \frac{X_q}{2} \nonumber\\
         & = & {\mathbb I}_n + \frac {X_q}{2!} + \frac {X_q^2}{3!}
               + \frac {X_q^3}{4!} + \ldots
\label{matrixfcn1}
\eea
and
\bea
  F(- X_q )^{-1} & = & e^{X_q}F(X_q )^{-1}
 = \frac {e^{\frac {X_q}2}}{\Sinch \frac{X_q}{2}} \nonumber\\
    & = & {\mathbb I}_n + \frac {X_q}{2} + \sum_{k\geq 1}
        (-1)^{k-1} \frac {B_k X_q^{2k}}{(2k)!},
\label{matrixfcn2}
\eea
where the $B_k$ are the Bernoulli numbers, $B_1 = \frac 16 , \;
B_2 = \frac 1{30} , \; B_3 = \frac 1{42}, \; B_4 = \frac 1{30}$,
etc., and generally,
$$ B_k = \frac {(2k)!}{\pi^{2k}\; 2^{2k-1}}\sum_{n=1}^{\infty}
     \frac 1{n^{2k}} \;\ \ \ k=1,2,3,....$$

   Later we shall need to express the Haar measure $\d\mu_G$
in terms of the coordinates of the Lie algebra, using the exponential
map. Writing
\be
  \d\mu_G (e^X) = m_G (\vec x_q , \vec x_p ) \;\d\vec x_q\;\d\vec x_p ,
\label{lefthaar5}
\en
the density function $m_G(\vec x_q , \vec x_p )$ is easily calculated,
using (\ref{expmap1}). Indeed, from (\ref{lefthaar1}),
\bea
\d\mu_G (e^X) & = &
\d \mu_G(e^{\frac {X_q}{2}}\Sinch\frac{X_q}{2}\vec x_p, e^{X_q}) \nonumber\\
& = &\frac{1}{|\det e^{X_q}|}\;|\det(e^{\frac
{X_q}{2}}\Sinch\frac{X_q}{2})|\;\d \mu_H(e^{X_q}) \;\d \vec x_p
   \nonumber \\
  & = & |\det(e^{\frac {-X_q}{2}}\Sinch\frac{X_q}{2})|\;
       \d \mu_H(e^{X_q})\;\d \vec
x_p
\label{lefthaar6}
\eea
It is also possible to write $\d \mu_H(e^{X_q})$ in terms of the
Lebesgue measure, $\d\vec x_q =\d x_1 \d x_2...\d x_n$,
times some density function $m_H$ \cite{Kirill-94},
\bea
 \d \mu_H(e^{X_q})& = & m_H (\vec x_q )\;d\vec x_q \nonumber\\
   & = & \le|\det\frac{1-e^{-adX_q}}{adX_q}\ri|
\;\d\vec x_q = \le|\det(e^{-ad\frac {X_q}{2}}
  \Sinch \le(ad\frac{X_q}{2}\ri))\ri|\;\d \vec x_q \; ,
\label{lefthaar7}
\eea
where $adX$ is the linear map on $\bfrakg$ which is the
infinitesimal generator of the adjoint
action $\Ad{g}, \;\; g \in G$:
\be
       adX(L)=[X,L], \quad \mbox{\rm and} \quad \Ad{g} = \Ad{(e^X)}
            = e^{adX}.
\label{infad1}
\en
Finally, the left Haar measure on G takes the form :
\be
\d \mu_G (e^X)=\le|\det(e^{\frac
{-X_q}{2}}\Sinch \le (\frac{X_q}{2}\ri ))
\;\det(e^{-ad\frac{X_q}{2}}\Sinch
\le (ad\frac{X_q}{2}\ri ))\ri|\;\d \vec x_q \;\d\vec x_p ,
\label {lefthaar8}
\ee
and the density function $m_G(\vec x)$ appearing in
(\ref{lefthaar5}) is:
\bea
m_G(\vec x) & = & |\det(e^{-\frac
{X_q}{2}}\Sinch \le (\frac{X_q}{2}\ri ))
\; \det(e^{-ad\frac{X_q}{2}}\Sinch
\le (ad\frac{X_q}{2}\ri ))| \nonumber\\
  & = & \vert\det F (-X_q )\;\det F(- adX_q)\vert
\label{lefthaar9}
\eea
\section{Orbits and invariant measures}\label{sec:orbinvmeas}
  It is now possible to determine the non-trivial coadjoint
orbits of $G$, which will be the main focus of our
attention.  These are orbits of fixed vectors in $\bfrakg^*$
under the coadjoint action (\ref{coadaction2}). Consider first the
vector $(\vec 0^{\;T} , \vec k^{\;T} ) \in \R^{2n},
\;\; \vec k \neq \vec 0$ and let
${\O^*}_{(\vec 0^{\;T} ,\vec k^{\;T} )}$ be its orbit under the
coadjoint action, i.e.,
\be
  {\O^*}_{(\vec 0^{\;T} ,\vec k^{\;T} )} = \{ (\vec \gamma_q^{\;T} ,
    \vec \gamma_p^{\;T} ) = (\vec 0^{\;T} ,\vec k^{\;T} )M(-\bh^{-1}\vec b ,
     \bh^{-1} ) \;\vert\; (\vec b , \bh ) \in \R^n \rtimes H \} .
\label{coadorbit1}
\en
Then, from (\ref{coadaction2}),
\bea
   \vec \gamma_q^{\; T}  & = &
   \vec k^{\; T}\bh^{-1}[\bfrakX\vec b] =
  \vec \gamma_p^{\; T}[\bfrakX\vec b], \nonumber\\
  \vec \gamma_p^{\; T} & = & \vec k^{\; T}\bh^{-1} .
\label{coadorbit2}
\eea
The vectors $\vec \gamma_p^{\; T}$ generate the orbit
$\widehat \O_{\vec k^{\;T}}$ of the subgroup $H$ in $\widehat{\R}$. We now
show that, for any $\vec \gamma_p^{\; T}$, the vector
$\vec \gamma_q^{\; T} = \vec \gamma_p^{\; T}[\bfrakX\vec b]$
can be identified with an element of the cotangent space of $\widehat \O_{\vec k^{\;T}}$ at this
point. Indeed, for any $i = 1,2, \ldots , n$,
consider a curve, $\vec u^{\;i}(t)^T$ in
$\widehat \O_{\vec k^{\;T}}$ of the type,
\be
    \vec u^{\;i} (t)^T = \vec \gamma_p^{\;T}e^{L^i t}, \qquad
    t \in [-\epsilon , \epsilon ] \subset \R .
\label{curve1}
\en
Then, $\vec u^{\;i} (0)^T = \vec \gamma_p^{\;T}$, and
\be
   \frac {d\vec u^{\;i} (t)^T}{dt}\Bigm|_{t = 0} =
                   \vec \gamma_p^{\;T}L^{i} := \vec t_p^{\;i\;T},
\label{tangentvect1}
\en
is a vector tangent to $\widehat \O_{\vec k^{\;T}}$ at
$\vec \gamma_p^{\;T}$. Recall that we are assuming that the action
of $H$ on $\widehat \R$ is open free. Hence the stability subgroup of the vector
$\vec k^{\;T}$ under the action  $\vec k^{\;T}
\mapsto \vec k^{\;T}\bh^{-1}$
is just the unit element of $H$ and  the orbit
$\widehat \O_{\vec k^{\;T}}$ is
an open set of $\widehat \R^n$, consequently of dimension $n$.
This implies
that the vectors $\vec t_p^{\;i\;T}$ are non-zero and
linearly independent and hence form a basis for the tangent space
$T_{\vec \gamma_p^{\;T}} \widehat{\O}_{\vec k^{\;T}}$ at
$\vec \gamma_p^{\;T}$. Let
$\vec t_p^{\;i\;T} = (t^{i1} , t^{i2} , \ldots , t^{in} )$,
in components, and define the matrix
\be
   \bT (\vec \gamma_p^{\;T}) = [t^{ij}] =
[\vec \theta_1 , \vec \theta_2 , \ldots, \vec \theta_n],
\label{tanmatrix}
\en
where the vectors $\vec \theta_i$ are its columns:
\be
  \vec \theta_i =\pmatrix{t^{1i}\cr t^{2i}\cr\ldots\cr t^{ni}},
\qquad
       i = 1,2, \ldots , n .
\label{cotanvect1}
\en
The vectors $\vec \theta_i$ form a basis for the cotangent
space $T^*_{\vec \gamma_p^{\;T}}\widehat \O_{\vec k^{\;T}}$
of $\widehat \O_{\vec k^{\;T}}$  at $\vec \gamma_p^{\;T}$.
Thus, if $b^i$ are
the components of the vector $\vec b$, we have
\be
   \vec \gamma_p^{\;T}[\bfrakX\vec b ] =
\sum_{i=1}^{n}b^i \vec \theta_i^{\;T} ,
\label{cotanvect2}
\en
implying that $[\vec \gamma_p^{\;T}[\bfrakX\vec b ]]^T
= [\bfrakX\vec b ]^T \vec \gamma_p$ is just a cotangent
vector at $\vec \gamma_p^{\;T}$.  Letting $\vec b$ run through all of $\R^n$,
these vectors generate the whole cotangent space at $\vec \gamma_p^{\;T}$.
Thus,
\be
   \O^*_{(\vec 0^{\;T} , \vec k^{\;T})} = T^*\widehat \O_{\vec k^{\;T}},
\label{coadorbit4}
\en
and if $\vec k^{\;T}$ is a vector such that the orbit $\widehat\O_{\vec k^{\;T}}$ is
open and free, the orbit $\O^*_{(\vec 0^{\;T} , \vec k^{\;T})}$ has dimension
$2n$. It is known \cite{BT-96} that if one such open free orbit exists,
then there exists a finite discrete set of them, corresponding to vectors
$\vec k_{j}^{\;T}, \;\; j = 1,2, \ldots , N < \infty$, for which
$\cup_{j = 1}^{N}\widehat\O_{\vec k_{j}^{\;T}}$ is dense in $\widehat \R^n$.\\
Similarly, let us compute the coadjoint orbit of a vector
$(\vec x^{\;T}, \vec 0^{\;T}) \in \widehat \R^{2n}$. As before,
\be
  \O^*_{(\vec x^{\;T}, \vec 0^{\;T} )} = \{ (\vec \gamma_q^{\;T} ,
    \vec \gamma_p^{\;T} ) = (\vec x^{\;T}, \vec 0^{\;T} )M(-\bh^{-1}\vec b ,
     \bh^{-1} ) \;\vert\; (\vec b , \bh ) \in \R^n \rtimes H \} ,
\label{coadorbit5}
\en
and again  from (\ref{coadaction2}),
\bea
   \vec \gamma_q^{\; T}  & = & \vec x^{\; T}M(\bh^{-1}) ,\nonumber\\
  \vec \gamma_p^{\; T} & = & \vec 0^{\; T}.
\label{coadorbit6}
\eea
and these orbits all have dimension lower than $2n$. From the point of view
of representation theory, these are the trivial orbits.
Using the coordinates $\gamma^i$ to identify $\bfrakg^*$ with
$\widehat \R^{2n}$, we arrive at the result:
\betheo\label{th:foliation}
If the action of $H$ on $\widehat \R^n$ is open free, the set of non-trivial
coadjoint orbits in $\bfrakg^*$ is finite and discrete and their union is dense
in $\bfrakg^*$.  Moreover, each nontrivial coadjoint orbit,
$\O^*_{(\vec 0^{\;T}, \vec k_{j}^{\;T})}$, is the cotangent
bundle, $T^* \widehat\O_{\vec k_{j}^{\;T}}$, of an open free orbit,
$\widehat\O_{\vec k_{j}^{\;T}}
\subset \widehat \R^n$, of a vector $\vec k_{j}^{\;T} \in
\widehat \R^n$ under the action of $H$. Under the coadjoint
action of $G = \R^n \rtimes H$, the dual space of its Lie algebra decomposes as
\be
  \bfrakg^* \;\; \simeq \;\; \widehat \R^{2n} \;\; = \;\;
     \left[\cup_{j = 1}^{N}\O^*_{(\vec 0^{\;T}, \vec k_{j}^{\;T})}
      \right] \; \cup \; V \;\; = \;\;
     \left[ \cup_{j = 1}^{N}T^*\widehat \O_{\vec k_{j}^{\;T}}\right] \;\cup \; V,
\label{coadorbit7}
\en
where $V$ is a set consisting of lower (than $2n$) dimensional
orbits and therefore
of Lebesgue measure zero in $\widehat\R^{2n}$.
\enth

   The orbits $\O^*_{(\vec 0^{\;T}, \vec k_{j}^{\;T})}$,
being {\em homogeneous symplectic manifolds\/} \cite{Kirill-76},
carry invariant measures under the coadjoint action
(\ref{coadaction1}) - (\ref{coadaction2}).
Indeed, if $d\vec \gamma^{\;T}$
denotes the Lebesgue measure $d\gamma^1\; d\gamma^2 \; \ldots \;
d\gamma^{2n}$, restricted to the orbit
$\O^*_{(\vec 0^{\;T}, \vec k_{j}^{\;T})}$,
then using (\ref{coadaction2}) and (\ref{modfcns}) it is
easy to check that under the coadjoint action
it transforms as,
\be
  d\vec \gamma^{\;'\;T} = \Delta_{G}(\vec b , \bh )
   d \vec \gamma^{\;T} .
\label{invmeas1}
\en
On the other hand, the mapping
$\kappa_j : \widehat\O_{\vec k_{j}^{\;T}} \rightarrow H$,
\be
   \kappa_j (\vec \gamma_p^{\;T} ) = \bh , \quad \mbox{\rm where}
   \quad  \vec \gamma_p^{\;T}  = \vec k_{j}^{\;T}\bh^{-1},
\label{orbhom1}
\en
is a homeomorphism. It is then straightforward to see that the measure
\be
   \d \Omega_{j}(\vec \gamma_q^{\;T} ,
     \vec \gamma_p^{\;T}) =  \sigma_j (\vec \gamma^{\;T} )^{-1}
\;\d\vec\gamma^{\;T} ,
     \qquad \sigma_j (\vec \gamma^{\;T} ) =
  \frac {\Delta_H [\kappa_j
    (\vec \gamma_p^{\;T} )]}{\vert\mbox{\rm det}\;[\kappa_j (\vec
 \gamma_p^{\;T} )]\vert} ,
\label{invmeas2}
\en
is invariant on $\O^*_{(\vec 0^{\;T}, \vec k_{j}^{\;T})}$
under the coadjoint action.

  Note, finally, that each one of the orbits
$\O^*_{(\vec 0^{\;T}, \vec k_{j}^{\;T})}$ is homeomorphic to
the group $G$ itself. Indeed, using (\ref{coadorbit2}),
(\ref{cotanvect2}) and (\ref{orbhom1}) let us define a map,
\be
  \widetilde{\kappa}_j :
  \O^*_{(\vec 0^{\;T}, \vec k_{j}^{\;T})} \longrightarrow
   \R^n \rtimes H , \qquad \widetilde{\kappa}_j (\vec \gamma_q^{\;T},
    \; \vec \gamma_p^{\;T}) = (\vec b , \bh ) =
   (\bT (\vec \gamma_p^{\;T})^{-1}
\vec \gamma_q , \; \kappa_j (\vec \gamma_p^{\;T})),
\label{orbhom2}
\en
where $\bT (\vec \gamma_p^{\;T})$ is the matrix of
tangent vectors defined in
(\ref{tanmatrix}). Then,
$\widetilde{\kappa}_j$ is a homeomorphism and it is straightforward
to verify that
\be
   \widetilde{\kappa}_j \circ \coAd{g_0} = L_{g_0}\circ
                                \widetilde{\kappa}_j ,
\label{adlintertwin1}
\en
where $L_{g_0}(g) = g_0 g, \;\; g \in G$. More explicitly, if
$(\vec \gamma_q^T , \vec \gamma_p^T ) \mapsto g = (\vec b , \bh )$
under $\widetilde{\kappa}_j$, then
\be
   \coAd{g_0}(\vec \gamma_q^T , \vec \gamma_p^T )
    \longmapsto (\vec b_0 , \bh_0 )(\vec b , \bh )
   = (\vec b_0 + \bh \vec b , \bh_0 \bh ).
\label{adlintertwin2}
\en
In other words, the homeomorphism $\widetilde{\kappa}_j$, from the
coadjoint orbit
$\O^*_{(\vec 0^{\;T}, \vec k_{j}^{\;T})}$ to the group $\R^n
\rtimes H$, intertwines the coadjoint action on the orbit
with the left action on the group and furthermore, under this
homeomorphism the invariant measure $\d \Omega_j$ on the orbit
transforms to the left Haar measure $\d \mu_G$ on the group.

   Before leaving this section, we describe a second,
in a way more intrinsic,
method for arriving at the invariant measure (\ref{invmeas2}),
using the fact that the orbits
$\O^*_{(\vec 0^{\;T}, \vec k_{j}^{\;T})}$ are symplectic
manifolds and thus carry $G$-invariant two-forms \cite{Kirill-76}
which can be computed using the structure constants of the group.
As before, $\{L^i\}_{i=1}^{2n}$ will be a basis for the Lie algebra
$\bfrakg$ and $\{L_i^*\}_{i=1}^{2n}$ the dual basis of $\bfrakg^*$.
The Lie algebra of the group $G$ is determined by the commutation
relations,
\be
  [L^i\;, \; L^j ] = \sum_{k=1}^{2n} c^{ij}_k\;L^k ,
\label{strconsts1}
\en
where the $c^{ij}_k$ are the structure constants. Thus, in
this basis, the linear map $adL^i$ has the matrix elements
$[adL^i]^j_k = c^{ij}_k$. Let $X^* = \sum_{i=1}^{2n}\gamma^i L_i^*
\in \O^*_{(\vec 0^{\;T}, \vec k_{j}^{\;T})}\subset \bfrakg^*$
and let us define a matrix
$\bTheta(\vec \gamma^{\;T})$ at this point by
\be
   [\bTheta(\vec \gamma^{\;T} )]^{ij} =
\sum_{k=1}^{n}[adL^i]^j_k \;\gamma^k
     = \sum_{k=1}^{n} c^{ij}_k \;\gamma^k .
\label{strconsts2}
\en
Using $\bTheta(\vec \gamma^{\;T} )$ matrix, we can now identify the Lie algebra
$\bfrakg$ with the tangent space,
$T_{X^*}\O^*_{(\vec 0^{\;T}, \vec k_{j}^{\;T})}$,
to the orbit $\O^*_{(\vec 0^{\;T}, \vec k_{j}^{\;T})}$
at the point $X^*$. (Note that this tangent space is naturally
isomorphic to $\bfrakg^*$ itself). Since
the orbit $\widehat\O_{\vec k^{\;T}_j}$ is open free, it has dimension $n$
and its cotangent bundle, i.e.,
the orbit $\O^*_{(\vec 0^{\;T}, \vec k_{j}^{\;T})}$,
has dimension
$2n$. Thus, $T_{X^*}\O^*_{(\vec 0^{\;T}, \vec k_{j}^{\;T})}$
has dimension $2n$ and in it we shall use the standard basis
$\{\frac {\partial}{\partial \gamma^i}\}_{i=1}^{2n}$.
Similarly, we shall use the dual
basis $\{\d\gamma^i \}_{i=1}^{2n}$ for the cotangent space,
$T_{X^*}^{*}\O^*_{(\vec 0^{\;T}, \vec k_{j}^{\;T})}$. From
(\ref{strconsts2}) we see that for $i = 1,2, \ldots , 2n$,
the vectors
$\sum_{j=1}^{2n} [\bTheta(\vec \gamma^{\;T} )]^{ij}
\frac {\partial}{\partial \gamma^j}$
form a linearly independent set of tangent vectors to the orbit
$\O^*_{(\vec 0^{\;T}, \vec k_{j}^{\;T})}$ at the point
$X^*$ (under the coadjoint action). Thus,
for $X = \sum_{i=1}^{2n}x_i L^i
\in \bfrakg$, it follows
that $\sum_{j=1}^n [\bTheta(\vec \gamma^{\;T} )\vec x ]^j\;
\frac {\partial}{\partial \gamma^j}$
defines a vector in
$T_{X^*}\O^*_{(\vec 0^{\;T}, \vec k_{j}^{\;T})}$ and hence we
have the identification map, $\phi_{X^*} :\bfrakg \longrightarrow
T_{X^*}\O^*_{(\vec 0^{\;T}, \vec k_{j}^{\;T})}$,
\be
\phi_{X^*}(X) = \sum_{i,j,k = 1}^{2n}c^{ij}_k \;x_i\;\gamma^k\;
 \frac{\partial}{\partial \gamma^j}
=  \sum_{i,j = 1}^{2n} [\bTheta(\vec \gamma^{\;T} )]^{ij}\;x_i \;
\frac{\partial}{\partial \gamma^j},
\label{identif1}
\en
as an isomorphism of vector spaces.
The $G$-invariant 2-form (symplectic form) is then defined as:
\be
\omega_{X^*}(\phi_{X^*}(X),\phi_{X^*}(L)) = <X^*;[X,L]> ,
\label{invform1}
\ee
which using (\ref{strconsts1}) and (\ref{identif1}) can be
expressed in the form,
\be
\omega_{X^*} = \sum_{i,j = 1}^{2n}
[\omega_{X^*}]_{ij}\; \d\gamma^i \wedge \d \gamma^j =
 \sum_{i,j = 1}^{2n}[\bTheta(\vec \gamma^{\;T} )]_{ij}\;
\d\gamma^j \wedge \d \gamma^i ,
\label{invform2}
\ee
where $[\bTheta(\vec \gamma^{\;T} )]_{ij}$ are the elements of the
inverse matrix $[\bTheta(\vec \gamma^{\;T} )]^{-1}$.
From this the $G$-invariant measure on the orbit
$\O^*_{(\vec 0^{\;T}, \vec k_{j}^{\;T})}$ is computed to be
$$
 \d\Omega_{j}(\vec \gamma^{\;T} )= \lambda
(\det [\omega_{X^*}])^{\frac12}\;\d \gamma^1\;
\d\gamma^{2}\;\ldots\;\d \gamma^{2n}
=\frac{\lambda}{(\det[\bTheta(\vec \gamma^{\;T})])^{\frac 12}}
\;\d \gamma^1\;\d\gamma^2 \;\ldots \;\d \gamma^{2n} ,
$$
where $\lambda$ is a constant. By multiplying the basis vectors
$L^i$ by appropriate constants, $\lambda$  can be made equal to
one. We shall assume that this has been done and then write,
\be
 \d\Omega_{j}(\vec \gamma^{\;T} )=
(\det [\omega_{X^*}])^{\frac12}\;\d \gamma^1\;
\d\gamma^{2}\;\ldots\;\d \gamma^{2n}
=\frac{1}{(\det[\bTheta(\vec \gamma^{\;T})])^{\frac 12}}
\;\d \gamma^1\;\d\gamma^2 \;\ldots \;\d \gamma^{2n} .
\label{strconsinvmeas}
\en
Comparing with (\ref{invmeas2}) we find,
\be
  \sigma_j (\vec \gamma^{\;T}) =
  (\det[\bTheta(\vec \gamma^{\;T})])^{\frac 12} .
\label{strconsdens}
\en
\section{Representations of $G$}\label{sec:repG}
In order to construct Wigner functions for the group $G = \G$
we shall use its quasi-regular representation. This
representation acts via the unitary operators
$U(\vec b , \bh )$ on the Hilbert space $\hil =
L^{2}({\R}^{n},\d \vec x)$:
\be
(U(\vec b,{\bh})f)(\vec x)=|\det {\bh}|^{-\frac{1}{2}}
f({\bh}^{-1}(\vec x -\vec b)) \qquad f \in \hil .
\label{31}
\ee
This representation is in general not irreducible, but is always
multiplicity free. Moreover, the existence of open free orbits
implies that every non-trivial irreducible sub-representation
of $G$ is contained in $U$ and each such representation is
{\em square integrable} \cite{BT-96} in a sense
to be made precise presently.\par
 In order to obtain the irreducible sub-representations of $U$,
it is useful to look at the unitarily equivalent representation
$\widehat{U} (\vec b , \bh ) = {\cal F}U(\vec b , \bh ){\cal F}^{-1}$,
where ${\cal F}:L^{2}({\R}^{n},\d \vec x)
\rightarrow L^{2}({\widehat \R}^{n},\d \vec k^{\;T})$
is the Fourier transform operator:
$$  ({\cal F}f)(\vec k^{\;T}) = \frac 1{(2\pi)^{\frac n2}}
    \int_{\R^n}e^{i\vec k^{\;T}\vec x}f(\vec x ) \;\d\vec x . $$
The action of $\widehat{U} (\vec b , \bh )$ on
a vector $\widehat f  \in \widehat\hil
= L^{2}({\widehat \R}^{n},\d \vec k^{\;T})$ is easily seen to have the
form,
\be
(\widehat U (\vec b,{\bh})\widehat f)(\vec k^{\;T})=
   |\det {\bh}|^\frac{1}{2}\;
e^{i\vec k^{\;T} \vec b}\; \widehat f (\vec k^{\;T} {\bh}) .
\label{32}
\ee
We shall also need the form of this representation, written in
terms of Lie algebra variables, using the exponential map
(\ref{expmap1}):
\be
(\widehat U (e^{-X})\widehat f) (\vec k^{\;T})=
 |\det [e^{-X_q}]|^{\frac 12}\; e^{(i \vec k^{\;T}
  F(-X_q ) \vec x_p)}\;
\widehat f (\vec k^{\;T}e^{-X_q}) .
\label{reprofalg}
\ee
 Let $\vec k^{\;T}_j \in \widehat \R^n , \;\; j =1,2,
\ldots , N$, be a
maximal set of vectors whose orbits $\widehat \O_{\vec k^{\;T}_j}$ under $H$
are open free and mutually disjoint. Then by
Theorem \ref{th:foliation}, $\cup_{j=1}^{N}\widehat \O_{\vec k^{\;T}_j}$ is
dense in $\widehat \R^n$ and $\cup_{j=1}^{N}T^{*}\widehat \O_{\vec k^{\;T}_j}$
is dense in the dual, $\bfrakg^*$, of the Lie algebra of $G$. Set
$\widehat \hil_j = L^2 (\widehat \O_{\vec k^{\;T}_j} , \d \vec k^{\;T} )$ (the
restriction of the Lebesgue measure to the orbit is implied). Then,
it is not hard to see that each of these spaces is an invariant
subspace for $\widehat U$. Moreover, the restriction
$\widehat U_j$, of $\widehat U$ to $\widehat \hil_j$, is irreducible
\cite{BT-96}, and is in fact the representation of $\G$
which is induced
from the character $\chi_j (\vec x) = \exp (i\vec k^{\;T}_j \vec x )$
of the abelian subgroup $\R^n$. Thus,
\be
  \widehat \hil = \oplus_{j=1}^N \widehat \hil_j, \qquad
  \widehat U (\vec b , \bh ) = \oplus_{j=1}^N
     \widehat U_j (\vec b , \bh ) ,
\label{Uhatdecomp}
\en
and it follows from Mackey's theory of induced representations
\cite{Mackey} for
semidirect product groups that these irreducible representations
exhaust all nontrivial irreducible representations of $G$.
\section{Square integrability of representations}
\label{sec:sq-integ}
The irreducible representations $\widehat U_j$
in (\ref{Uhatdecomp}) all have one
other property, of importance to us here. These representations are
{\em square integrable\/} \cite{AAG-2000}.
Recall that a unitary irreducible representation $U$ of a group
$G$ on a Hilbert space $\hil$ is square integrable if
there exists a non-zero
vector $\eta \in \hil$, called an  {\em admissible vector\/},
such that:
\be
c(\eta)=\int_{G}|<U(g) \eta|\eta>|^2\d \mu_G (g)< \infty
\label{29}
\ee
The existence of one such vector and irreducibility of the
representation imply that the set of all admissible vectors
${\cal A}$ is dense in $\hil$. If the group is unimodular
then $\cal A$ coincides with $\hil$, otherwise it is a
proper subset of it. (For a more detailed
description of square integrable representations and their
properties, see e.g., \cite{AAG-2000}).
For any square integrable representation $U$ there exists a unique
positive operator $C$ on $\hil$ whose domain coincides
with ${\cal A}$ and
such that if $\eta_1, \eta_2 \in {\cal A}$ and
$\phi_1,\phi_2 \in \hil$ the following {\em orthogonality
relation} holds:
\be
\int_{G}\overline{<U(g)\eta_2|\phi_2>}<U(g)\eta_1|\phi_1> \d \mu_G
 \; \; = \;\; <C \eta_1|C
\eta_2><\phi_2|\phi_1>
\label{orthogonreln1}
\ee
This result is due to Duflo and Moore \cite{Du-Mo}
and the operator $C$ is usually referred to in the literature
as the Duflo-Moore operator. If $G$ is unimodular, $C$ is a multiple
of the identity, otherwise, it is an unbounded operator.\par
   For semidirect product groups $G = \R^n \rtimes H$ of the type
discussed in the previous sections, with open free orbits, the
irreducible representations $\widehat{U}_j (\vec b , \bh )$,
appearing in the decomposition (\ref{Uhatdecomp}) are
all square integrable  and one has the result \cite{BT-96}:
\bt\label{c-function}
Let $H$ be a closed subgroup of $GL_n(\R)$ and let $G=\R ^n \rtimes
H$.
Let $\widehat\O_{\vec k^{\;T}_j}$ be an open free $H$-orbit in
$\widehat \R^n$.
Then the restriction $\widehat{U}_j (\vec b , \bh )$, of the
quasiregular representation to the
Hilbert space $L^2 (\widehat \O , \d\vec k^{\;T})$, is irreducible and
square integrable. The corresponding
Duflo-Moore operator $C_j$ assumes the form:
\be
 (C_j f)(\vec k^{\;T}) = (2\pi)^\frac{n}{2} [c_j
             (\vec k^{\;T} )]^\frac{1}{2}f(\vec k^{\;T}),
\label{33}
\ee
on $L^{2}(\widehat\O_{\vec k^{\;T}_j}  , \d \vec k^{\;T})$,
where $c_j :\widehat \O_{\vec k^{\;T}_j} \longrightarrow \R ^+$ is
a positive, Lebesgue measurable function
which transforms under the action of $H$ as:
\be
c_j(\vec k^{\;T} {\bh})=\frac{\Delta_H(\bh)}{|\det{\bh}|}
c_j(\vec k^{\;T}),
\label{34}
\ee
for almost all $\vec k^{\;T}$ (with respect to the Lebesgue
measure).
  Furthermore, every irreducible representation of $G$ is of this
type and the quasi-regular representation is a
multiplicity-free direct sum of these representations.
\et
    It has also been shown in \cite{BT-96} that
$c_j(\vec k^{\;T})$ is precisely the density function
which converts the Lebesgue measure $\d \vec k^{\;T}$, restricted to
the orbit $\widehat \O_{\vec k^{\;T}_j}$, to the invariant measure
$\d\nu_j$ on it:
\be
  \d\nu_j (\vec k^{\;T}) = c_j(\vec k^{\;T})\; \d\vec k^{\;T}
\quad \mbox{\rm and} \quad \d\nu_j (\vec k^{\;T}\bh )
    = \d\nu_j (\vec k^{\;T}),
\label{invorbmeas1}
\en
and can be defined simply to be the transform of the left Haar
measure $\d\mu_H$ of $H$ under the homeomorphism (\ref{orbhom1}),
\be
 \d\nu_j (\vec k^{\;T} ) = \d\mu_H (\kappa_j (\vec k^{\;T})).
\label{invorbmeas2}
\en
    If $\vec \gamma_p^{\;T} \in \widehat\O_{\vec k^{\;T}_j}$ is an
arbitrary point and $\vec \gamma_p^{\;T} = \vec k^{\;T}_j \bh^{-1}$,
(see (\ref{orbhom1})), then in view of (\ref{34}) we
may set,
$$
c_j(\vec \gamma^{\;T}_p )=\lambda\;\frac{|\det{\bh}|}{\Delta_H(\bh)},
$$
for almost all $\vec \gamma^{\;T}_p \in \widehat \O_{\vec k^{\;T}_j}$
(with respect to the Lebesgue measure), where $\lambda$ is a
constant. (Clearly, with this choice of of the density
$c_j(\vec \gamma^{\;T}_p )$ the invariance
condition in (\ref{invorbmeas1}) is satisfied). In view of
(\ref{invorbmeas2}) we may, by multiplying $\d\mu_H$ by a
constant if necessary, make $\lambda = 1$. Assuming that this has
been done,
we may write (for almost all $\vec \gamma^{\;T}_p$),
\be
c_j(\vec \gamma^{\;T}_p )=
  \frac {|\det{[\kappa_j (\vec \gamma^{\;T}_p )]}|}
  {\Delta_H [\kappa_j (\vec \gamma^{\;T}_p )]},
\label{duflmoorop3}
\ee
Comparing with (\ref{righthaar1}),(\ref{invmeas2}) and
(\ref{strconsdens}), and using the
homeomorphism $\widetilde \kappa_j :
 \O^*_{(\vec 0^{\;T}, \vec k^{\;T}_j )}
 \rightarrow \R^n \rtimes H$ in (\ref{orbhom2}), we have the
result,

\bt
Let $H$ be a closed subgroup of $GL_n(\R)$ and let
$G=\R ^n \rtimes H$.
Let $\widehat \O_{\vec k^{\;T}_j}$ be an open free $H$-orbit in
$\widehat \R^n$ and let $T^{*}\widehat\O_{\vec k^{\;T}_j}$
be its cotangent bundle with invariant measure $d\Omega_j$.
Then the following equalities hold:
\be
   c_j(\vec \gamma^{\;T}_p )^{-1} = \sigma_j (\vec \gamma^{\;T}) =
  (\det[\bTheta(\vec \gamma^{\;T})])^{\frac 12} =
  \;\Delta_G [\widetilde \kappa_j (\vec \gamma^{\;T})],
\label{duflmoorop4}
\en
(except at most on a set of measure zero),
where $c_j$ is the function defining the Duflo-Moore
operator of the unitary irreducible
representation $\widehat{U}_j$ of $G$, associated
to the orbit $\widehat\O_{\vec k^{\;T}_j},\;\;
\sigma_j(\vec \gamma^{\;T})$ is the Radon-Nikodym derivative,
$\d\vec \gamma^{\;T}/d\Omega_j$, at the point $\vec \gamma^{\;T} =
(\vec \gamma_q^{\;T},\vec \gamma_p^{\;T})= (\gamma^1 , \gamma^2,
\ldots , \gamma^n , \gamma^{n+1}, \ldots , \gamma^{2n}) \in
T^{*}\widehat\O_{\vec k^{\;T}_j}, \;\;
[\bTheta(\vec \gamma^{\;T})]^{ij}=
\sum_{k = 1}^{2n}c^{ij}_k \gamma^k , \;\;
c^{ij}_k$ being the structure constants of $G$, and
$\widetilde{\kappa}_j$ the homeomorphism between
$T^{*}\widehat\O_{\vec k^{\;T}_j}$ and
$\R ^n \rtimes H$, normalized so that
$\widetilde{\kappa}_j (\vec 0^{\;T}, \vec k^{\;T}_j )
= (\vec 0 , e)$.
\et
\section{Construction of general Wigner functions}\label{sec:cons-Wig-fcns}
  It is known (see, for example, \cite{AAG-2000})
that the orthogonality relations (\ref{orthogonreln1}) for
a square integrable representation $U$ of
the group $G$ have an
extension to Hilbert-Schmidt operators on $\hil$. Let
${\cal B}_2 (\hil )$ denote the Hilbert space of all Hilbert-Schmidt
operators $\rho$ on $\hil$. This Hilbert  space is equipped with the
scalar product,
$$
 \langle \rho_1 \;\vert\; \rho_2 \rangle_{\cal B} =
    \mbox{\rm Tr}[\rho_1^* \rho_2 ] .
$$
Then there exists a dense set ${\cal D} \subset {\cal B}_2 (\hil )$
such that for any $\rho \in {\cal D}$, the (closure of) the operator
$U(g)^* \rho C^{-1}$ is of trace class ($C$ being the Duflo-Moore
operator). Furthermore, the function
\be
  f_{\rho} (g) = \mbox{\rm Tr}[U(g)^* \rho C^{-1}],
\label{Wig_trans1}
\en
is an element of $L^2 (G, \d\mu_G )$ and moreover,
\be
   \Vert f_{\rho} \Vert_{L^2 (G, \d\mu_G )}^2 =
   \Vert \rho \Vert_{\cal B}^2 .
\label{Wig_trans2}
\en
Thus, we may define an isometric linear map,
$\widetilde{\mathfrak W}
: {\cal B}_2 (\hil ) \longrightarrow L^2 (G, \d\mu_G )$ which, for
$\rho \in {\cal D}$, is given by (\ref{Wig_trans1}) and is then
extended by continuity to all of ${\cal B}_2 (\hil )$.

  We are now ready to give the definition of the general Wigner function.
However, it is first necessary to make an additional assumption on the group $G$,
that there exist a symmetric subset $N_0$ of Lie algebra $\bfrakg$,
such that the exponential map
restricted to it is a bijection onto a dense set (in $G$), the
complement of which has Haar measure zero: $\mu (G-\exp(N_0))=0$. The Wigner
function is then defined as a Fourier-like transform of (\ref{Wig_trans1}):
\be
W(\rho\;|\vec \gamma^T)=\frac{1}{(2\pi)^\frac{n}{2}}
\int_{N_0}e^{-i\vec \gamma^T \vec x }{\rm Tr}[U(e^{-X}\rho C^{-1}] [\sigma
(\vec \gamma^T)m_G (\vec x)]^\frac{1}{2}\d \vec x
\ee
or equivalently, if $\rho = \vert\phi\rangle\langle\psi\vert$, 
\be
W(\phi ,\psi\;|\; \vec \gamma^T)=\frac{1}{(2\pi)^\frac{n}{2}}
\int_{N_0}e^{-i\vec \gamma^T \vec x }<C^{-1}\psi|U(e^{-X})\phi>[\sigma
(\vec \gamma^T)m_G (\vec x)]^\frac{1}{2}\d \vec x
\label{Wignerfcn1}
\ee
In this expression, 
 $\vec \gamma^T \in \bfrakg ,\ \phi \in\hil $ and $\psi$
is in the range
of the Duflo-Moore operator $C$ related to the representation $U$.
The density function $m_G$ again expresses the Haar measure on
$G$ in terms of the Lebesgue measure $\d \vec x$ on $\bfrakg$
(see (\ref{lefthaar5})):
\be
\d \mu_G (e^X)= m_G(\vec x) \d \vec x
\label{16}
\ee
where $X \in \bfrakg$ is expressed in terms of the
components of the vector $\vec x$ in the basis
$\{L^1,L^2,...L^{2n}\}$ (i.e., $X=\sum_{i=1}^{2n} x_iL^i$).
The function $\sigma$ is defined by expressing the Lebesgue measure $\d X^*$
in $\bfrakg ^*$ in terms of invariant measures $\d\Omega_\lambda$ on the
coadjoint orbits $\O_\lambda ^* $ as follows:
\be
\d X^*=\d \kappa(\lambda) \sigma_\lambda (X^*)\d \Omega_\lambda(X^*)
\label{measuredecomp1}
\ee
where the index $\lambda$ parametrizes coadjoint orbit and $\d\kappa(\lambda)$ is a measure on the parameter space.
This decomposition is not guaranteed in general and has to be
assumed or proved for specific cases.
\section{Basic properties of general Wigner function}
The appearance  of $ \sigma_\lambda$ in the formula for the Wigner
function is necessary in order to have the following important
covariance property :
\be
W(U(g)\rho U(g)^*|\vec \gamma^T)=W(\rho|{Ad_{g^{-1}}}^{\#}\vec
\gamma^T)
\label{cov2}
\ee
which clearly can be regarded as a generalization of the covariance
(\ref{covariance}) for the original Wigner Function.
The overlap condition (\ref{overlap1}) in this more general setting becomes:
\be
\int_{\bfrakg^*}\overline{W(\rho_1|\vec \gamma^T)}W(\rho_2|\vec
\gamma^T)[\sigma(\vec \gamma)^T]^{-1}\d \vec \gamma^T={\rm Tr}[\rho^*_1\rho_2]
\label{overlap2}
\ee
As expected, the general Wigner function does not enjoy all the properties
of the original function $W^{QM}$. For example, not both marginal properties
(\ref{marginal1}), (\ref{marginal2}) are generally satisfied, or can be given a 
natural meaning, in view of their dependence on a special choice of coordiantes 
for the phase space over which the function is defined. Additionally, the domain 
of the general Wigner function could span more than a single coadjoint orbit, and 
hence more than a single physical phase space. We shall return to these points 
later, in the context of 
semidirect product groups of the type described in section (\ref{sec:mathprelims}).
\section{General Wigner function for semidirect pro\-duct groups}
Before proceeding to the case of semidirect product groups,
few comments about the general construction are in order.
The original Wigner function, 
presented in Section \ref{sec:stanwigfcn} has the same form as the general
Wigner function, with both density functions $m(\vec x)$ and $\sigma(\vec \gamma^T)$ 
being equal to one and the Duflo-Moore operator being a multiple of identity operator.
There is a more subtle difference, however: The original Wigner function arises 
from an irreducible representation of the Heisenberg-Weyl group $G_{HW}$, and these 
representations 
are only square integrable with respect to the homogeneous space $G_{HW}/\Theta$
($\Theta$ being the phase group), 
and not with respect to the whole group. (On the other hand, as shown in 
\cite{AFK-2001}, this additional generality is only an apprent one, in 
this case, and is easily subsumed in a more general theory, based on the 
Plancherel transform.) It is nevertheless still worth stressing here,
that the general general procedure for constructing
Wigner functions presented in this paper does rest on two requirements: 
first, the decomposability
(\ref{measuredecomp1}) of the Lebesgue measure $\d X^*$ and second, 
that we consider only groups with square integrable representations.

Semidirect product groups, with open free orbits, which we consider here, satisfy both
these conditions. Using their square integrable
representations, (\ref{Uhatdecomp}), we can rewrite the general Wigner function
(\ref{reprofalg}) as:
\bea
W(\widehat \phi,\widehat \psi|\vec \gamma^T)
& = &
\frac{1}{(2\pi)^n}\int_{N_0}\d
\vec x e^{-i\vec \gamma^T \vec x} \int_{\O ^*}\d \vec \omega^T
\overline{C^{-1}\widehat \psi (\vec \omega^T)}|\det\ e^{-X_q}|^{\frac{1}{2}}
\cr
&\times &\exp(i\vec \omega^T e^{-\frac{X_q}{2}}\Sinch\ \frac{X_q}{2}
\vec x_p)\widehat \phi (\vec \omega^T e^{-X_q})
[\sigma(\vec \gamma^T)m(\vec x)]^{\frac{1}{2}}
\label{GWF1}
\eea

Changing variables: $\vec {\omega}'\;^T= \vec \omega^T e^{-\frac{X_q}{2}}\Sinch\
\frac{X_q}{2}$ and using the form for the density function $m(\vec x)$
given in Eq.(\ref{lefthaar9}) we obtain:
\bea
W(\widehat \phi,\widehat \psi|\vec
\gamma^T)& = & \frac{1}{(2\pi)^n}\int_{N_{0q}}\int_{\R ^2}\d \vec x_q \d \vec x_p
 e^{-i\vec \gamma^T_q \vec x_q}\cr
&\times &
\int_{\O ^*}\d \vec \omega'^T e^{i(\vec {\omega'^T}-\vec \gamma^T_p)\vec
x_p}
\overline{C^{-1}\widehat \psi
\le (\vec \omega'^T
\frac{ e^{\frac{X_q}{2}}}{\Sinch\ \frac{X_q}{2}}\ri)}\widehat \phi
\le(\vec \omega'\frac{ e^{-\frac{X_q}{2}}}{\Sinch\
\frac{X_q}{2}}\ri)\cr
&\times &
\sigma(\vec \gamma^T)^{\frac{1}{2}} \le|
\det\frac{ e^{-\frac{X_q}{2}}}{\Sinch\
\frac{X_q}{2}}\ri|^{\frac{1}{2}}
 \le| \det\le ( e^{-ad\frac{X_q}{2}}\Sinch\ ad\frac{X_q}{2}\ri)\ri|^{\frac{1}{2}}
\label{GWF2}
\eea
We have shown in Theorem \ref{c-function} that the
Duflo-Moore operator in this case is related to the decomposition
(\ref{measuredecomp1}) of Lebesgue measure in $\bfrakg$ and 
is expressible in terms of the 
structure constants of the Lie algebra $\bfrakg$.
Applying ($\ref{duflmoorop4}$) together with ($\ref{34}$) and
integrating over $\vec x_p$ we finally obtain:
\bea
W(\widehat \phi,\widehat \psi|\vec \gamma^T) & = & \int_{N_{0q}}\d
\vec x_q e^{-i\vec \gamma^T_q \vec x_q}\overline {\widehat \psi \le(\vec
\gamma^T_p\frac{ e^{\frac{X_q}{2}}}{\Sinch\ \frac{X_q}{2}}\ri)}\widehat\phi
\le(\vec\gamma_p\frac{ e^{-\frac{X_q}{2}}}{\Sinch\ \frac{X_q}{2}}\ri)\cr
&\times &
c\le(\vec\gamma^T_p\frac{1}{\Sinch\ \frac{X_q}{2}}\ri)^{-\frac{1}{2}}
c(\vec\gamma^T_p)^{-\frac{1}{2}}\le|\frac{\det(\Sinch\ ad\frac{X_q}{2})}
{\det( \Sinch\ \frac{X_q}{2})}\ri|^{\frac{1}{2}}
\label{Wignerfinalform}
\eea
Here we used the fact that the domain $N_0$ of the exponential map $ \exp :\bfrakg
\rightarrow G $ in the case of semidirect product groups, is given by $N_{0q}
\times \R^n$, where $N_{0q}$ is the corresponding domain
the exponential map $\exp :\h \rightarrow H$. Again we assume that this map
is a bijection onto a dense set in $H$ such that its complement has measure zero.

   It is easily seen that if we integrate the 
Wigner function (\ref{Wignerfinalform}) with respect to the measure 
$c(\vec \gamma_p^T) \d \vec \gamma_q^T$, we obtain a marginal property of the form:
\be
\int_{\R^n}W(\widehat \phi,\widehat \psi|\vec \gamma^T) c(\vec
\gamma_p^T)\d \gamma_q^T =\overline{\widehat \phi(\vec \gamma_p^T)}\widehat 
\psi(\vec \gamma_p^T)
\ee
as a generalization of (\ref{marginal1}). Unfortunately, the second 
property (\ref{marginal2}) does not have a simple form any more.

\section{Domain of the Wigner Function}
As mentioned earlier, the advantage of using the original
Wigner function is  that it allows us to represent a quantum state
or a signal (in the case of signal analysis)
as a function on a phase space (position-momentum or time-frequency).
Thus, ideally, we would like our Wigner function to be 
supported on a single coadjoint orbit, (which together with its 
symplectic form $\omega_{X^*}$ 
in (\ref{invform2}) can be considered as a phase space).
We investigate  now, under what conditions
this is the case for the Wigner functions just derived.

Recall first that the open free coadjoint orbits $\O^*_i$ in $\bfrakg^*$,
for semidirect product groups, are in one to one correspondence with
open free $H$-orbits $\widehat \O_i \subset \widehat \R^n$: indeed
according to Theorem \ref{th:foliation}, any $\O^*_i$ is a cotangent bundle of the form
 $\O^*_i=T^* \widehat \O_i=\widehat \O_i\times \R^n$.
Let $W_{\widehat \O_i}$ denote the Wigner function derived from
a representation of $G$ acting on $\hil = L^2(\widehat \O_i  )$,
which can be conveniently be thought of as the closed subspace of
$L^2(\widehat \R^n , d\vec x )$ of functions which vanish almost everywhere outside
$\widehat \O_i$.
We are going to find sufficient conditions for the Wigner function
$W_{\widehat \O_i}$ to have support concentrated on the corresponding coadjoint orbit
 $ \O^*_i=\widehat \O_i \times \R^n $.

Let us start by introducing a polynomial function $\Delta $,
\be
\Delta(\vec\gamma^T)=\det\pmatrix{\vec \gamma^T L_1\cr \vec \gamma^T L_2\cr
...\cr \vec \gamma^T L_n},
\label{deltaf}
\ee
where $\{L_1,..,L_{n-1},L_n\}$ is a basis in $ \bfrakg$.
It is very convenient to discuss the orbit structure in $\widehat \R^n$ with
the use of this function $\Delta$. A point $\vec \gamma^T$ belongs to the
open (free) orbit if and only if it satisfies $\Delta(x) \neq 0$ and
a point belongs to an orbit of dimension $< n$ if and only if $\Delta(\vec \gamma^T)=0$.
We have the following:
\beprop
\label{domainofWf}
Let $G$ be a semidirect product group $\R^n \rtimes H$, such that  $H$ acts
on $\widehat \R^n$ with open, free orbits $\{\widehat \O_i\}_{i=1}^m$.
If an orbit $\widehat \O_i$ is a  dihedral cones (i.e.
if the zero level set of the function $\Delta$ in (\ref{deltaf}), restricted to it, can be
decomposed into hyperplanes) then the Wigner function $W_ {\widehat \O_i}$ has 
support concentrated on
the corresponding coadjoint orbit $ \O^*_i=\R^n \times \widehat \O_i$.
\enprop
To prove it we will need the following lemma:
\belem
\label{invhyperplane}
If a hyperplane $\Pi(\vec \gamma ^T)=0$ is a subset of $\Delta(\vec
\gamma ^T)=0$ then it is invariant under $H$.
\enlem
{\bf Proof of Lemma \ref{invhyperplane}}
We will show first that we can always find $\vec \gamma_0 ^T \in
\Pi^{-1}(0)$ such that there exists a neighborhood $U_{\vec \gamma_0^T}$
satisfying
\be
U_{\vec \gamma_0^T} \cap \Pi^{-1}(0)=U_{\vec \gamma_0^T} \cap \Delta ^{-1}(0)
\label{hyperplane1}
\ee
Let us introduce a basis  $ \{Z_1,...,Z_n\}$ in $\widehat \R^n $ such that 
the first $n-1$ elements constitute a basis in the hyperplane $\Pi(\vec \gamma^T)=0$. 
In these coordinates, the the hyperplane can be written as 
$\Pi(\vec \gamma^T)=\gamma^n$ and the function $\Delta$ can be factored as:
\be
\Delta(\gamma^T)=(\gamma^n)^k P(\vec \gamma^T)
\label{deltafactors}
\ee
such that $P(\vec \gamma)$ does not contain $\gamma^n$ as a factor.
Then $P(\gamma^1,...\gamma^{n-1},0)\equiv 0$ iff $P(\vec \gamma)\equiv
0$ (which would imply $ \Delta(\vec \gamma)\equiv 0$, a
contradiction) or $P(\vec \gamma)$
 contains $ \gamma^n $ as a factor, which would contradict (\ref{deltafactors}).
Thus we can always choose $\vec \gamma_0=(\gamma_1,...,\gamma_{n-1},0)$ 
such that $P(\vec \gamma_0 )=r\not=0$ and $\Pi(\vec \gamma_0)=0$.
Since $P$ is a polynomial, there exists an open neighborhood 
$U_{\vec\gamma_0^T}$ of $\gamma_0^T$ such that 
$P(\vec \gamma_0^T) \in (r-\epsilon, r+\epsilon)$.
Thus we have (\ref{hyperplane1}).\par
Now, for any $\vec \gamma^T \in U_{\vec \gamma_0^T}$
 the intersection of its orbit $\O_{\vec \gamma ^T}$ with $U_{\vec \gamma_0^T}$ belongs to $\Pi$.
This implies that for every $\vec \gamma^T\in U_{\vec \gamma_0^T}$,
$\bfrakh \vec\gamma^T  \in \Pi$. We can choose a basis of $\Pi$ 
formed by $N-1$ linearly independent elements $\{ Z'_1,...,Z'_{N-1}\}\subset U_{\vec \gamma_0^T}$.
Since $\bfrakh Z'_i \subset \Pi$ is true for every basis element
then also for every $\vec \gamma^T \in \Pi \ ,\bfrakh \vec \gamma^T
 \subset \Pi$, i.e. the hyperplane $\Pi$ is stable under $\bfrakh$ and
 hence also under $H$ (by exponentiation).
{\bf QED}

{\bf Proof of Proposition \ref{domainofWf}}
One sees from (\ref{Wignerfinalform}) that a sufficient condition
for the Wigner transform to preserve the decomposition into orbits $\O^*_i$
is that the point $\vec
\gamma^T\frac{e^{\frac{X_q}{2}}}{\Sinch\frac{X_q}{2}}$ does not leave
$\widehat \O_i$ as $X_q$ varies in $N_{0q}$, or equivalently that the
`$\Sinch$' map preserves the orbits (which is not guaranteed because
$\Sinch (X)$ is not an element of the group $H$).
Let us take again a basis $\{Z_1,...,Z_{n-1},Z_n\}$
 in $\widehat \R^n$ such as the first $n-1$ elements
belong to the $(n-1)$-dim hyperplane as in Lemma \ref{invhyperplane}.
In the coordinates introduced above an element $X$ of the Lie  algebra
$\bfrakh$ of the group $H$ is of the form:
$$
X=\pmatrix{X_{1,1\ \ }&...&X_{1,n-1}&0\cr
...\cr X_{n-1,1}&...&
X_{n-1,n-1}&0\cr X_{n,1\ \ }&...&X_{n,n-1}&X_{n,n}}
$$
because $X$ must preserve the hyperplane $\gamma^n=0$.
Calculating the $\Sinch$ of such element $X$ we obtain:
$$
S=\Sinch (X)=\pmatrix{S_{1,1}\ \ &...&S_{1,n-1}\ &0\cr...\cr
S_{n-1,1}&...&S_{n-1,n-1}&0\cr S_{n,1}\ \ &...&S_{n,n-1}\
&\Sinch(X_{n,n})}
$$
Notice that $\Sinch(X_{n,n})>0$ from definition (\ref{sinchfcn1}).
Applying $\Sinch(X)$ to any vector $\vec \gamma^T$ in $\widehat \R^n$ written in
the basis $\{Z_i\}^n_1 $ we have :
$$
\Sinch(X)(\gamma^1,...,\gamma^{n-1},\gamma^n)=
({\gamma^1 }\pr,...,{\gamma^{n-1}}\pr,\Sinch(X_{n,n}))\gamma^n)
$$
Therefore the sign of $\gamma^n$ remains unchanged, which also means
that the hyperplane $\gamma^n=0$ divides $\widehat \R^n$ into two
halfspaces, invariant under the $\Sinch $ map.
Since $\Delta^{-1}(0)$ is a union of hyperplanes $\Pi_1 \cup \Pi_2
...\cup \Pi_r$ we can repeat the argument for each of them, proving
that each open orbit is preserved.
{\bf QED}
\section{Examples}
It is now an easy task to explicitly compute Wigner functions for
particular cases of groups  from its general form for semidirect product
groups, (\ref{Wignerfinalform}). Let us consider first
examples of connected $4$-dimensional
semidirect product groups $ G= \R^2 \rtimes H$ with open free $H$-orbits
in $\widehat\R^2$, i.e. when $H$ is diagonal group, SIM(2) or
one of the infinite family of $H_c$ groups. We work out the  Wigner
functions for all such groups.
Next we present an interesting example of an $8$-dimensional group
$G=H\rtimes H^*$, where $H$ is a vector space of quaternions and $H^*$
a group of invertible quaternions. The Wigner functions constructed here 
are all candidates for use in image analysis in various dimensions (see
also, \cite{AACW-1999}, \cite{AKM-2000} and \cite{BB-92}).  

\subsection{ The diagonal group}

Let $G=\R^2\rtimes H$ where $H$ is the diagonal subgroup of $GL_2(\R)$ that is
$H=\{\pmatrix{ a_1 &0\cr 0&a_2\cr} \; ; a_1,a_2\in \R-\{0\}\}$. 
The Wigner functions are defined on the coadjoint $G-$orbits, $\O
^*_{\vec\gamma _{ij}^T}$, of the elements $\vec\gamma_{ij}^T =(0,0,i,j),
i=\pm 1, j=\pm 1$ the union of which is
dense in ${\widehat \R}^4$.
\bea
W(\widehat \phi,\widehat \psi|\vec \gamma^T)& = & \frac{1}{2\pi} \int_{\R^2}\d
 x_1 \d x_2 e^{-i \gamma^1 x_1-i \gamma^2 x_2}\overline {\widehat \psi (
\gamma_1\frac{ e^{\frac{x_1}{2}}}{\Sinch\ \frac{x_1}{2}})}\cr
&\times &
\widehat\phi
(\gamma_2\frac{ e^{-\frac{x_2}{2}}}{\Sinch\
\frac{x_2}{2}})\frac{|\gamma^3\gamma^4|}{\Sinch\frac{x_1}{2}\Sinch\frac{x_2}{2}},
\label{46}
\eea
where an element of $\h$ has the form $ X_q=\pmatrix{x_1&0\cr 0&x_2}$
and the corresponding element in $H$ is $e^{X_q}=\pmatrix{e^{x_1}&0\cr
0&e^{x_2}}$.
We have also used the following relations:
$$\Sinch\ \frac{X_q}2 =\pmatrix{\Sinch\frac{x_1}2&0\cr
0&\Sinch\frac{x_2}2}$$
$$c(\vec \gamma_p^T)=|\gamma^3 \gamma^4 | . $$
\subsection{The SIM($2$) group}
Let $G$ denote the $SIM(2)$ i.e. the group of dilations rotations and
translations in $\R^2$. $G=\R^2\rtimes H$ where
$H=\{\pmatrix{a&-b\cr b&a}:(a,b)\in\R^2-\{0,0\}\}$.
The Wigner function is defined on $\O_{\vec
\gamma_0^T}^*=\{(\gamma^1,\gamma^2,\gamma^3,\gamma^4)
:\ (\gamma_3,\gamma_4)\not=(0,0)\}$ ( the coadjoint orbit of $\vec
\gamma_0=(0,0,1,0)$ ).
This case was studied extensively in \cite{AKM-2000}. This time,
however we can make use of the Wigner function for semidirect product
groups to obtain the recult immediately.   \\
Denoting the element of the lie algebra $\h$ as $X_q=\pmatrix{
\lambda&-\theta\cr\theta&\lambda}, \theta \in
(0,2\pi),\ \lambda \geq 0$ and the corresponding element of the group
$H$ as $e^{X_q}=\pmatrix{e^\lambda \cos \theta&-e^\lambda \sin
\theta \cr e^\lambda \sin \theta&e^\lambda \cos \theta}$
the corresponding Wigner function is :
\bea
W(\widehat \phi,\widehat \psi|\vec \gamma^T)& = & 
   \frac{(\gamma^3)^2+(\gamma^4)^2}{2\pi}\int_{N_0q}
 e^{-i\gamma_1 \lambda-i\gamma_2\theta}\overline {\widehat \psi (\vec
\gamma_p\frac{ e^{\frac{X_q}{2}}}{\Sinch\ \frac{X_q}{2}})}\widehat\phi
(\vec\gamma_p\frac{ e^{-\frac{X_q}{2}}}{\Sinch\ \frac{X_q}{2}})\cr
&\times &
\frac{\lambda^2+\theta^2}{2 \cosh \lambda -2\cos\theta}\d\lambda\d\theta ,
\eea
We have used the following relations:
$$\det\
\Sinch \frac{X_q}2=\frac{2\cosh \lambda -2\cos \theta}{\lambda^2+\theta^2}$$
and$$c(\vec \gamma_p^T)=|(\gamma^3)^2+(\gamma^4)^2|^{-1}$$
Since $H$ is abelian, $\det\ (\Sinch\ ad\frac{X_q}2)=1$.
\subsection{The one parameter family of groups $H_c$}
Consider now the one parameter family of groups $H_c=\{\pmatrix{a&0\cr
b&a^c}\ :\ a,b\in\R \ , a>0\} $ for $c\not=0$.
The Wigner functions are defined on coadjoint orbits 
$ \O_+=\{(\gamma^1,\gamma^2 ,\gamma^3,\gamma^4
\ :\gamma^4>0\}$ and $\O_-=\{(\gamma^1,\gamma^2 ,\gamma^3,\gamma^4\
:\gamma^4<0\}$.\\
An element of the Lie algebra $\h _c$ is 
$X_q=\pmatrix{x_1&0\cr x_2&cx_1}$, and the corresponding element of a group
$H_c$ is
$e^{X_q}=\pmatrix{e^{x_1}&0\cr
\frac{x_2}{(c-1)x_1}(e^{x_1c}-e^{x_1})&e^{x_1c}}$
(in the case c=1 we should take the $\lim_{c\rightarrow 1}$ )\\
The Wigner function takes the form:
\bea
W(\widehat \phi,\widehat \psi|\vec \gamma^T) & = & \frac{(|\gamma^4|^2)}{2\pi}\int_{N_0q}
 e^{-i\gamma^1 x_1-i\gamma^2 x_2}\overline {\widehat \psi (\vec
\gamma_p\frac{ e^{\frac{X_q}{2}}}{\Sinch\ \frac{X_q}{2}})}\widehat\phi
(\vec\gamma_p\frac{ e^{-\frac{X_q}{2}}}{\Sinch\ \frac{X_q}{2}})\cr
&\times &
\frac 1{\Sinch\frac{x_1 c}2}(\frac{\Sinch\frac{x_1 (c-1)}2}{\Sinch \frac{x_1}2
\Sinch\frac{x_1 c}2})^\frac 12\; dx_1\;dx_2 , 
\eea
where we have used the following relations:
$$\Sinch \frac{X_q}2=\pmatrix{\Sinch\frac{x_1}2&0\cr \frac
1{1-c}\frac{x_2}{x_1}(\Sinch\frac{x_1}2-\Sinch\frac{cx_1}2)&\Sinch\frac{cx_1}2}$$
$$\det( \Sinch\ ad\frac{X_q}2)=\Sinch\frac{(c-1)x_1}2$$
and $$c(\vec \gamma_p)=|\gamma^4|^{-2}$$.
\subsection{Quaternionic groups}
Quaternions constitute a (nonabelian) field of numbers; they can be
thought of as an extension of the complex numbers, similar to the way that complex
numbers are an extension of the real ones. More specifically they are obtained by adding
two more ``imaginary units'', customarily denoted by $\mathbf j,\mathbf k$, such that 
the following relations are
fulfilled:
\be
\mathbf j^2 = \mathbf i^2 =\mathbf  k^2 = -1;\ \mathbf i\mathbf j = \mathbf k,\
\mathbf j\mathbf k = \mathbf i,\ \mathbf k\mathbf i= \mathbf j.
\ee
The generic quaternion can be written as $x_0 + x_1\mathbf i + x_2 \mathbf j + x_3\mathbf k$  
where $x_0,x_1,x_2,x_3$ are real numbers, or as
$ z_0 + z_1{\mathbf j}$, where $z_0 = x_0 + x_1{\mathbf i} $ and $z_1 = x_2+x_3{\mathbf i}$  
are complex numbers. 
A very practical way of dealing with quaternions is to represent them as $2\times 2$ matrices with
complex entries
\be
q := \pmatrix{x_0+ix_1&x_2+ix_3\cr -x_2+ix_3&x_0-ix_1}\ .
\ee
By identifying the set of quaternions with $\R^4$ one can endow this latter space with a notion of multiplication.
It is also worthwhile to recall that any nonzero quaternion admits a (multiplicative) inverse which can be expressed by
taking the inverse of  the matrix representing it. \par
Let us consider now the semidirect product group $ G=H \rtimes H^* $
where $H$ denotes the vector space of quaternions and $H^*$ the group
of invertible quaternions.
An element of the group can be written in the form:
$$g=\pmatrix{h_q & h_p \cr {0}&{1}}$$ where $ h_q \in H^*$ and
$h_p \in H$.
Then an element of the lie algebra $\bfrakg= Lie(G)$ is:
$$X=\pmatrix{X_q&X_p\cr { 0}&{ 0}}$$
Where $X_q$ and $X_p$ are both quaternions which can be written in
coordinates as :
$$
X_q=\pmatrix{x_0+ix_1&x_2+ix_3\cr -x_2+ix_3&x_0-ix_1}\ \ \ \
 X_p=\pmatrix{x_4+ix_5&x_6+ix_7\cr -x_6+ix_7&x_4-ix_5}
$$
The group can be equivalently written, in a manner more consistent
with the rest of this paper, as $\R^4 \rtimes M(h_q)$ where $M(h_q)\in
GL(4,\R)$ is of the form :
$$M(h_q)=\pmatrix{x_0&-x_1&-x_2&-x_3\cr x_1&x_0&-x_3&x_2\cr
x_2&x_3&x_0&-x_1\cr x_3&-x_2&x_1&x_0}$$
The quaternionic notation makes it easy to relate this group to the
 $G_1=\R \rtimes \R^*$ and $G_2=\C \rtimes \C^*=SIM(2)$,
which are the wavelet groups in 1 and 2 dimensions respectively.
It seems quite natural to use the field of quaternions to define
a wavelet group in 4 dimensions. The concept of wavelet groups can be
therefore extended (in rather straightforward way) to any Clifford
algebra.
The Wigner function is defined on the single coadjoint orbit $\O ^* =\bfrakg ^* -\{0\}$.
\bea
W(\widehat \phi,\widehat \psi|\vec X^*)& = & \frac{|X_p^*|^4}{(2\pi)^4}\int_{\R ^4}
 e^{-i<X_q^*,X_q>}\overline {\widehat \psi (
X_p^*\frac{ e^{\frac{X_q}{2}}}{\Sinch\ \frac{X_q}{2}})}\widehat\phi
(X_p^*\frac{ e^{-\frac{X_q}{2}}}{\Sinch\ \frac{X_q}{2}})\cr
&\times &
\frac{1}{16}\frac{|X_q|^4}{ (\cosh^2\frac{x_0}{2} -\cos^2
\frac{R}{2})^2}\frac{\sin R}{R}\d X_q
\eea
where $R=(x_1^2+x_2^2+x_3^2)^\frac12$.
We also used,
$$c(X^*)=|X_p^*|^{-4}$$
$$\d \mu_G (e^X)=\det(e^{-\frac{X_q}2} \Sinch\frac{X_q}2)\frac{\sin
R}{R}\d X $$
All these computations can be easily repeated for any Clifford algebra.
\subsection{A group H which does not satisfy the assumption of Theorem
\ref{domainofWf}}
Consider a 3-dimensional group $H$, the Lie  algebra $\bfrakh$ of which 
is generated by the following elements:
\be
L=\pmatrix{1&0&0\cr 0&1&0\cr 0&0&1}\ \ F^1=\pmatrix{0&0&0\cr 1&0&0\cr 0&1&0}\
\ F^2=\pmatrix{1&0&0\cr 0&0&0\cr 0&0&-1}
\ee
The orbit structure in $\widehat \R^3$ is given by the equation:
$\Delta(\vec \omega^T)=-\frac13 \omega_3(-2\omega_3\omega_1+\omega_2^2)=0$ which clearly cannot be 
decomposed into hyperplanes.
We have the following open orbits in $\widehat \R^n$:

\noindent $\widehat \O_1$ - above the hyperplane $\omega_3=0$ and inside the cone
$-2\omega_3\omega_1+\omega_2^2<0\ \ (\Delta >0)$,

\noindent $\widehat \O_2$ - above the hyperplane $\omega_3=0$ and outside the cone
$-2\omega_3\omega_1+\omega_2^2>0\ \ (\Delta <0)$,

\noindent $\widehat \O_3$ - below the hyperplane $\omega_3=0$ and inside the cone
$-2\omega_3\omega_1+\omega_2^2< 0\ \ (\Delta>0)$,

\noindent $\widehat \O_4$ - below the hyperplane $\omega_3=0$ and outside the
cone($-2\omega_3\omega_1+\omega_2^2>0\ \ \Delta<0)$.

In order to see that the $\Sinch$ map does not preserve orbits let us
choose a point in $\widehat \O_2$ : $\vec
\omega_0^T=(\omega_1,\omega_2,\omega_3)$ and apply
to it $\Sinch(tF^1) = \pmatrix {1&0&0\cr 0&1&0\cr \frac 1 6 t^2 &0
&1}$, $t\in\R$. Then one can compute  $\Delta(\vec
\omega_0^T \Sinch(tF^1)) = \frac 1 9\,\omega_{{3}}\left
(6\,\omega_{{3}}\omega_{{1}}+{\omega_{{3}}}^{2}
{t}^{2}-3\,{\omega_{{2}}}^{2}\right) $.
It is clear that, as a function of $t$, it changes sign whenever
$2\,\omega_{{3}}\omega_{{1}}-\,{\omega_{{2}}}^{2}<0$.
 This also means that the $\Sinch $ map mixes two orbits
$\widehat \O_1$ with  $\widehat \O_2$ and also $\widehat \O_3$ with
$\widehat \O_4$. By a continuity argument, this mixing property holds
for a suitable open neighborhood of $F^1$ in the Lie algebra, i.e. a
set of positive Lebesgue measure.
As a  consequence, a Wigner function $W(\widehat \phi,\widehat
\psi|X^*)$ corresponding to two functions supported in $\widehat
\O_1$, $\widehat \phi, \widehat \psi \in L^2(\widehat \O_1)$
will have its support spread  on {\em both} coadjoint orbits
$\O^*_1$ and $\O^*_2$.
 To see that let us fix ${\vec \gamma_p}^T={\vec \omega_0}^T \in
\widehat \O _2$.
Then the  Wigner function,  as a function of ${\vec
\gamma_q}^T \in \widehat \R^n$, is just  the Fourier transform of a function
$F(X_q)$
\be
 W_{\omega_0^T}(\widehat \phi,\widehat \psi|{\vec \gamma_q}^T)=\int_{\R^n}\d
    \vec x_q e^{-i{\vec \gamma_q}^T \vec x_q}F(X_q)
\ee
 where:
\bea
 F(X_q)& = & \overline {\widehat \psi \le(\vec
         \omega_0^T\frac{ e^{\frac{X_q}{2}}}{\Sinch\
\frac{X_q}{2}}\ri)}\widehat\phi
          \le(\vec\omega_0^T\frac{ e^{-\frac{X_q}{2}}}{\Sinch\
\frac{X_q}{2}}\ri)\cr
          &\times & c\le(\vec\omega_0^T\frac{1}{\Sinch\ \frac{X_q}{2}}\ri)^{-\frac{1}{2}}
          c(\vec\omega_0^T)^{-\frac{1}{2}}\le|\frac{\det(\Sinch\ ad\frac{X_q}{2})}
          {\det( \Sinch\ \frac{X_q}{2})}\ri|^{\frac{1}{2}}\ .
\eea
Since the map $\Sinch(X_q)$ (and $\Sinch(X_q)^{-1}$ by the same argument) brings $\vec \omega_0^T$ from
$\widehat \O_2$ to $\widehat \O_1$ ( support of $\widehat \phi ,
\widehat \psi$) the function $F(X_q)$ is not identically zero,
e.g., for $X_q$ in a suitable open neighborhood of $F^1$.
 Then
its Fourier transform $W_{\omega_0^T}(\widehat \phi,\widehat
\psi|{\vec \gamma_q}^T)$ is also not identically zero. This means that
the Wigner function $W(\widehat \phi ,\widehat \psi|X^*)$ does not
vanishes outside the orbit $\O^*_1$.
\section*{Acknowledgments}
We should like to thank M. Bertola and H. F\"uhr for discussions and suggestions.  
We would also like to acknowledge
financial support from the Natural Sciences and Engineering Research Council, 
Canada and the Fonds pour
la formation de Chercheurs et l'Aide \`a la Recherche, Qu\'ebec.


\end{document}